\documentclass[12pt]{article}
\usepackage{amsmath}
\usepackage{amssymb}

\input tcilatex

\QQQ{Language}{
British English
}

\begin{document}

\title{All Universes Great and Small}
\author{John D. Barrow$^1$ and Hideo Kodama$^2$ \\
$^1$DAMTP, Centre for Mathematical Sciences, \\
Wilberforce Rd., Cambridge CB3 0WA, UK\\
email: J.D.Barrow@damtp.cam.ac.uk\\
$^2$Yukawa Institute for Theoretical Physics, \\
Kyoto University, Kyoto 606-8502, Japan\\
email: kodama@yukawa.kyoto-u.ac.jp}
\maketitle

\begin{abstract}
If the topology of the universe is compact we show how it  significantly
changes our assessment of the naturalness of the observed structure of the
universe and the likelihood of its present state of high isotropy and near
flatness arising from generic initial conditions.We also identify the most
general  cosmological models with compact space.
\end{abstract}

Interest has grown steadily in the possibility that the universe may be
finite in volume but not possess positive spatial curvature \cite{star}.
Textbook expositions of cosmology largely ignore this possibility, but there
are many reasons to contemplate the prospect of an open universe with a
compact topology \cite{top}. Many compact spaces of negative curvature can
possess unusually small volumes and actions, and formulations of quantum
cosmology require spatial finiteness in order for cosmological
wave-functions to exist. Spacetimes that tunnel out of 'nothing' may be more
likely to emerge with small compact spaces of negative curvature than as
closed universes. It is interesting that the curvature radius of a compact
open universe provides a characteristic length-scale to which any
topological scale of periodicity might be simply related. For the more
philosophically minded, the prospect of an infinite cosmological space may
be unattractive because any event occurring here and now with finite
probability must be occurring infinitely often elsewhere \cite{ellis}.
Observations of the expansion of the universe now imply that it will
continue expanding forever, and only a compact spatial topology can
reconcile this with a universe of finite volume. However, this is not the
only effect of a compact topology upon the expansion of the universe.
Topology is a global property that can exert a powerful local impact.
Indeed, we shall see that it can introduce a radically new element into our
attempts to understand the structure of the universe. 

One of the imperatives of modern gravitation research has been to reconcile
the observed properties of the universe with those that emerge from generic
initial conditions. The general cosmological solution of Einstein's
equations is specified by $4$ arbitrary $3$-dimensional functions of space
in vacuum and by $8$ such functions if a perfect fluid is present.
Nonetheless, the universe we observe is within one part in $10^5$ of a
special isotropic and spatially homogeneous (SH) solution with zero
curvature characterised by a \textit{single} parameter. This surprisingly
close proximity of our visible universe to a state that is isotropic and
homogeneous, and almost spatially flat, is difficult to understand if matter
has always obeyed conventional equations of state and exerted an attractive
gravitational stress. Under such circumstances we know that isotropic
expansion will not be the end-result of long-term expansion: isotropic
universes are unstable. Moreover, any perturbation from exact spatial
flatness will also grow as the expansion proceeds: to arrive at the present
state of near flatness after 14 billion years of expansion requires a
fantastic degree of initial fine tuning. A favourite resolution of these
paradoxes is to give up the assumption that matter was always
gravitationally attractive. If the density $\rho $ and pressure $p$, violate
the strong energy condition, so that $\rho +3p<0,$ for some sufficiently
long interval of time in the very early history of the universe, then the
expansion will temporarily accelerate and can be driven arbitrarily close to
an isotropic state of zero curvature. Such an interlude of accelerated
expansion is called 'inflation' \cite{guth}.

Remarkably, if an open universe possesses compact topology, many of these
problems are transformed and the present state of the universe appears
natural even in the absence of inflation. The simplest arena in which to
make a complete analysis is provided universes which are anisotropic but
spatially homogeneous. There are only nine such universes, first classified
by Bianchi as types $I-IX$ \cite{bian}, \cite{Taub}, and their general
solutions possess $4$ (or $8$) \textit{constants} in vacuum (or with perfect
fluid) content, respectively. The behaviour of these universes has only been
extensively studied when space is \textit{not} compact, \cite{CH}. In
non-compact space, a small neighbourhood of generic SH initial conditions
containing the open Friedmann universe evolves asymptotically towards a
family of anisotropic plane-wave spacetimes of Bianchi type $VII_h$, whilst
generic flat universes approach universes of Bianchi type $VII_0$ with small
expansion anisotropy and bounded curvature anisotropy. This scenario changes
dramatically if the topology of space is compact because compact topology
introduces severe restrictions on the anisotropies that can exist in
universes with zero or negative curvature \cite{BK}.

Anisotropic universes can be viewed as isotropic universes to which
particular gravitational waves are added. The periodic boundary conditions
that the waves must satisfy on compact negatively curved spaces are
extremely restrictive because their flow is chaotic. As a result, \textit{no
anisotropic open universes with compact spaces are possible at all when they
contain Friedmann universes as particular cases}. In particular, the Bianchi 
$VII_h$ anisotropies that cause the generic instability \cite{CH} of open
isotropic Friedmann universes with non-compact spaces can no longer exist.
Remarkably, in the presence of any anisotropy at all, flat compact universes
are more likely than open universes if they contain Friedmann universes as
special cases. The situation in non-compact flat and open universe which
inspired the invention of inflation is completely reversed. \textit{There is
no flatness problem.}

What are the most general SH universes when space has a compact topology? In
general, the addition of spatial compactness to open or flat spaces
introduces additional parameters (see Table 1). This is due to the
appearance of the moduli degrees of freedom and to a decrease in the freedom
of diffeomorphisms connecting physically equivalent solutions. For example,
in the case of a simple $3$-torus, we must specify the lattice in the
Euclidean $3$-space so as to define the $3$-torus in a rotationally
invariant way. We need $3$ parameters to specify the lengths of $3$ vectors
generating the lattice and $3$ parameters to specify the relative direction
angles of these vectors. Hence, adding their time derivatives, we need $12$
parameters. If we take into account the Hamiltonian constraint and the time
translation freedom, the total number reduces to $10$. By contrast, without
compactness the total is only $1.$

\begin{table}[tbp]
\begin{tabular}{ccccc}
Bianchi Type & \multicolumn{2}{c}{Vacuum} & \multicolumn{2}{c}{Perfect fluid}
\\ 
& Non-compact & Compact & Non-compact & Compact \\ 
Class A &  &  &  &  \\ 
$I$ & 1 & 10 & 2 & 11 \\ 
$II$ & 2 & 6 & 5 & 9 \\ 
$VI_0$ & 3 & 4 & 7 & 8 \\ 
$VII_0$ & 3 & 8 & 7 & 12 \\ 
$VIII$ & 4 & $4+N_m$ & 8 & $8+N_m$ \\ 
(LRS) & 2 & $2+N_m^{\prime}$ & 3 & $3+N_m^{\prime}$ \\ 
$IX$ & $-$ & 4 & $-$ & 8 \\ 
Class B &  &  &  &  \\ 
$III$ & 3 & $2+N_m^{\prime}$ & 7 & $3+N_m^{\prime}$ \\ 
$IV$ & 3 & --- & 7 & --- \\ 
$V$ & 1 & 0 & 5 & 1 \\ 
$VI_h$ & 4 & --- & 8 & --- \\ 
$VII_h$ & 4 & 0 & 8 & 1
\end{tabular}
\caption{Maximal degrees of freedom for spatially open and compact Bianchi
vacuum models and for those with perfect fluid.}
\label{tbl:count}
\end{table}

In Table 1 we see that when compactness is not assumed for open and flat
spaces, the four most general SH universes are those of Bianchi types $%
VI_h,VII_h,VIII$ and $IX$ \cite{ell}. When all spaces are compact, no
universes of type $VI_h$ can exist, whilst those of type $VII_h$ must be
isotropic and are no longer generic. Moreover, universes of type $IX$ which
contain the Friedmann closed universes of positive curvature now cease to be
among the most general. One significant feature of the parameter count in
the spatially compact open universes is the difference between the counts
for vacuum and perfect fluid models. Among the spatially compact \textit{%
vacuum } Bianchi models, the type $I$ model can be the most generic unlike
in the usual situation with non-compact topology, where it is the least
generic. In contrast, when a perfect fluid is present, the parameter count
for type $VII_0$ models is always larger than that for type $I$ models in
any given space topology. Therefore, the most general locally homogeneous 
\textit{perfect-fluid} spacetimes that include the flat isotropic model are
the  type $VII_0$ models in the spatially compact case just as they are in
the non-compact case. But when we come to ask what is the most general
universe of all, a new and unusual situation emerges. 

The Bianchi type $VIII$ universes become the most general when space is
compactified in a complicated way (see Table 1). In this case, the parameter
count increases over the non-compact case by the number of moduli degrees of
freedom, $N_m$. Since each compact space with the full type-$VIII$ spatial
anisotropy possesses a unique Seifert bundle structure over an orientable
compact orbifold covered by the hyperbolic plane, $N_m$ is expressed in
terms of the genus $g$ and the number, $k,$ of conic singularities of the
base orbifold, as 

\[
N_m=2k+6g-6.
\]

Hence the parameter count increases without bound as the topology becomes
more and more complicated. What is interesting for this type is that if we
restrict the model to the locally rotationally symmetric (LRS) case, the
moduli degrees of freedom increases to 
\[
N_m^{\prime }=N_m+2g
\]
for the same topology. This increase can beat the decrease in the dynamical
degrees of freedom describing the local geometry if $g>3$.\footnote{%
A similar phenomenon happens for Bianchi type $III$. In this case the
spacetime must be LRS if the space is compact, the parameter count is given
by the same number as that for the LRS case of type $VIII$. Thus it becomes
as generic as type $VIII$, although the non-compact counterpart is less
generic.} This result is intriguing, because Bianchi type $III$ or $VIII$
models do not contain Friedmann universes as special cases -- although they
can get arbitrarily close to them. This suggests that it is most probable
for a universe to have approximate LRS type $III$ or $VIII$ symmetry on
large scales if its space is compact and the observed region of the universe
is sufficiently flat. This will be directly reflected in the pattern of the
microwave background radiation on large angular scales.

We have found that topology has a major impact upon the possible forms that
expanding universes can take. Almost all previous investigations of compact
universes have assumed isotropic expansion. However, the introduction of any
finite anisotropy reveals the impact of topology: anisotropic curved spaces
are not easy to sustain. If their topology is required to be compact then
the possibilities become fewer still \cite{thur}. 

Many questions remain to be answered. Does the introduction of
inhomogeneities restrict the possible universes still further? How does the
possibility of additional dimensions of space change the constraints that
topology imposes upon deviations from perfectly isotropic expansion.? While
at present it is very difficult to prove strong results about inhomogeneous
spaces, we can expect an extension of the results in the SH\ type $VII_h$
system because the rigidity theorem used to derive them  was formulated in a
very general way. The theorems of Borel \cite{borel} and Mostow \cite{mostow}
imply that a broad class of compact spaces do not admit an anisotropic,
locally homogeneous metric in any dimension \cite{note}. This suggests a
powerful extension of our results to more general situations and in higher
dimensions. 

If space is compact then its high level of isotropy and proximity to
flatness becomes far less mysterious. Compact open universes are necessarily
isotropic when they are homogeneous; flat universes are more probable than
open or closed universes; and the most general universes that lie close to
isotropy and homogeneity are expected to possess a complex  topology whose
signature may have left an indelible imprint in the microwave background
radiation. Maybe a small world is simpler after all.


\begin{thebibliography}{99}
\bibitem{note}  Unfortunately, their physical implications to
higher-dimensional cosmology are not certain at present because the most
favourable compactifications in the string theory utilize Calabi-Yau
manifolds, which generally admit the moduli freedom of the complex and the
K\"{a}hler structures, to respect supersymmetry. However, since
supersymmetry must be broken and harmful moduli fields should be eliminated
in a final theory, there is a possibility that the rigidity theorem in
higher dimensions may play a significant role in finding the final theory.
\end{thebibliography}
\end{document}